\documentclass[journal,12pt,onecolumn]{IEEEtran}

\pdfoutput=1
\usepackage{textcomp}
\usepackage{stfloats}
\usepackage{verbatim}
\usepackage{balance}

\usepackage{amsmath,amsfonts}
\usepackage{amssymb,amsthm}
\usepackage{array}
\usepackage{algorithm}
\usepackage{algorithmic}
\usepackage{url}
\usepackage{cite,graphicx}
\usepackage{subfigure} 
\usepackage{citesort}
\usepackage{fancyhdr}
\usepackage{mdwmath}
\usepackage{mdwtab}
\usepackage{caption}

\newtheorem{theorem}{Theorem}

\newtheorem{lemma}{Lemma}

\newtheorem{corollary}{Corollary}

\makeatletter
\def\ScaleIfNeeded{%
\ifdim\Gin@nat@width>\linewidth \linewidth \else \Gin@nat@width
\fi } \makeatother
\usepackage{soul} 

\usepackage[dvipsnames]{xcolor}
\usepackage{color} 
\usepackage{multirow}
\usepackage{flafter}

\usepackage{lipsum} 
\usepackage{cuted}
\usepackage{flushend}

\usepackage{tcolorbox}
\tcbuselibrary{breakable, skins, theorems}
\usepackage{graphicx}
\usepackage{epstopdf}
\usepackage{pdfpages}
\usepackage{paracol}
\usepackage{booktabs}
\usepackage{framed}

\usepackage[justification=centering]{caption}
\usepackage{threeparttable}    
\usepackage{pifont}

\begin{document}
	\setlength{\lineskiplimit}{0pt}
	\setlength{\lineskip}{0pt}
	\setlength{\abovedisplayskip}{6pt}   
	\setlength{\belowdisplayskip}{6pt}
	\setlength{\abovedisplayshortskip}{6pt}
	\setlength{\belowdisplayshortskip}{6pt}	
    \title{
Caching Content Placement and Beamforming Co-design for IRS-Aided MIMO Systems with Imperfect CSI
    	}

\author{
Meng~Gao,
Yang~Wang, Huafu~Li,
Junqi Guo
\thanks{
	This work was supported in part by Science and Technology Project of Shenzhen under Grant JCYJ20200109113424990. 
    \emph{(Corresponding author: Yang Wang.)}
}
\thanks{Yang Wang, Meng Gao, Junqi Guo are with the School of Electronics and Information Engineering, Harbin Institute of Technology, Shenzhen 518055, China (email: yangw@hit.edu.cn, \{20b952022,19b952012\}@stu.hit.edu.cn).}
\thanks{
	Huafu Li is with the China Mobile Information Technology Co., Ltd., Shenzhen 518000, China (e-mail: lihuafu@chinamobile.com).
	}
}
\maketitle
\begin{abstract}
When offloading links encounter deep fading and obstruction, edge caching cannot fully enhance wireless network performance and improve the QoS of edge nodes, as it fails to effectively reduce backhaul burden.
The emerging technology of intelligent reflecting surfaces (IRS) compensates for this disadvantage by creating a smart and reconfigurable wireless environment. Subsequently, we jointly design content placement and active/passive beamforming to minimize network costs under imperfect channel state information (CSI) in the IRS-oriented edge caching system.
This minimization problem is decomposed into two subproblems.
The content placement subproblem is addressed by applying KKT optimality conditions. We then develop the alternating optimization method to resolve precoder and reflection beamforming. Specifically, we reduce transmission power by first fixing the phase shift, reducing the problem to a convex one relative to the precoder, which is solved through convex optimization. Next, we fix the precoder and resolve the resulting reflection beamforming problem using the penalty convex-concave procedure (CCP) method.
Results demonstrate that our proposed method outperforms uniform caching and random phase approaches in reducing transmission power and saving network costs. Eventually, the proposed approach offers potential improvements in the caching optimization and transmission robustness of wireless communication with imperfect CSI.
\end{abstract}
\vspace{-0.3cm}
\section{Introduction}

\IEEEPARstart{E}{dge} 
 caching can effectively alleviate the backhaul burden on the core network in  the industrial Internet of Things (IoT) system and enhance quality of service (QoS) at wireless edge nodes~\cite{9457078,li2022phase}. However, the benefits of edge caching cannot be fully realized when offloading links experience deep fading and obstruction by obstacles.
Further, these non-deterministic traffic (i.e., critical ambient monitoring, software update, test report uploads) cannot be easily accessed by the IoT systems due to communication interruptions or weak connections, which will seriously threaten the efficiency and productivity of the factory~\cite{9599656,zhang2023irs,9815179,10220203}.
The blockages and obstacles such as metal structures, machinery, or other densely deployed equipment in the factory environment will cause signal attenuation or shadowing~\cite{9599656,zhang2023irs}. 
Moreover, wireless channel is affected by path loss and multipath fading located in the complex electromagnetic space of the factory, which causes a degradation of signal quality and reducing the QoS.
Fortunately, unlike massive multiple input multiple output (MIMO) and relays (i.e., amplify-and-forward and decode-and-forward) with RF chains,
the near-passive and full duplex intelligent reflecting surfaces (IRS) $\footnote{There have been other terminologies introduced which share similarities with IRS (i.e., intelligent walls~\cite{wu2020beamforming,subrt2012intelligent}, passive intelligent mirror~\cite{huang2019reconfigurable,wu2021intelligent}),
reconfigurable intelligent surface~\cite{huang2019reconfigurable,pan2021reconfigurable,pan2022overview}
	 .}$ 
	 can repair damage caused by wireless propagation environments through establishing virtual line of sight links in a cost-effective manner, 
	 especially in cases of poor signal quality and insufficient transmission coverage at the edge, which has the ability to overcome high attenuation and ensure reliable communication~\cite{huang2019reconfigurable,wu2020towards,pan2021reconfigurable,pan2022overview,yuan2021reconfigurable,basar2019wireless,9110587}.
Therefore, it can effectively tackle the forthcoming demands and challenges encountered within IRS enabled edge caching toward IoT networks.

Furthermore, unlike caching in content delivery network, edge caching distributes content files to edges~\cite{9457078,li2022phase}. Mainly, the uncertainty of cache performance is often caused by imperfect content transmission links on wireless media. 
Then, the propagation environment may be affected by unpredictable user mobility and obstruction, increasing the difficulty of caching strategies and content delivery.
Caching designs inherently enhance performance from communications perspective due to their connection between communication and caching strategies.
To maximize the storage of hotspot content in service nodes (or minimize the distance between hotspot content and users) and improve the probability of successfully transmitting user-requested files, it is essential to design an effective caching placement strategy.

The works in~\cite{e2015cache,e2016edge} studied the impact of maximum popularity caching strategy on coverage or interruption probability in single-layer cellular networks and heterogeneous networks, respectively. ~\cite{d2016cache} investigated the impact of maximum popularity caching on spectral efficiency in heterogeneous networks. The studies in~\cite{e2015cache,e2016edge,d2016cache} have shown that caching can significantly improve the performance of networks. ~\cite{7435255} studied the trade-off between cache capacity and backhaul link bandwidth in single-layer cellular networks. 
Research indicates that as the cache capacity of a base station (BS) increases, the bandwidth demand on the backhaul link decreases.
 It is worth noting that, the storage space of BS is always limited in actual networks, so caching strategy will result in many popular content not being placed in the BS. Overall, the maximum popularity caching strategy causes all BS to store the same content, resulting in a decrease in file diversity and wasting the storage space of the BS. ~\cite{b2015optimal} first proposed the concept of random cache and designed a simple random storage document in a single-layer cellular network. The random geometry tools and convex optimization theory are used to analyze and optimize the successful transmission probability of typical users. ~\cite{wen2017cache} proved that the cache strategy optimization problem for heterogeneous networks with cache below 6GHz under non-uniform signal-to-interference ratio threshold is non convex, and obtained a suboptimal cache probability. Furthermore, the author proved that the cache probability optimization problem is convex under a uniform signal-to-interference ratio threshold, and obtained the optimal cache probability using a sequential calculation method. ~\cite{li2018optimization} derived a closed form expression for the probability of successful file transmission in cache heterogeneous networks below 6GHz and maximized the probability of successful transmission by optimizing cache probability.
 Meanwhile, the author studied the trade-off between BS density and cache size based on uniform caching strategy. ~\cite{8058445} proposed an effective collaborative caching strategy that minimizes the average file configuration cost. ~\cite{8625415} proposed a caching strategy based on truncated Zipf distribution to maximize the cache hit rate in D2D caching networks. ~\cite{8304636} analyzed the performances of a hybrid network consisting of small cellular base stations with caching capabilities and traditional small cellular base stations connected to the backhaul link. The author demonstrated that the maximum popularity caching strategy can maximize regional spectral efficiency when the slope of the Zipf distribution is high. 

Recently, a significant amount of research has been conducted to explore the potential of IRS in enhancing the reliability and robustness of wireless  networks~\cite{wu2020beamforming,wu2019intelligent,9933780,abeywickrama2020intelligent,saglam2022deep}.
Unlike approaches that consider continuous phase shifts at IRS reflecting elements~\cite{wu2019intelligent}, ~\cite{wu2020beamforming} aims to minimize transmit power by optimizing precoder and discrete reflection phase shifts under phase amplitude independence and perfect CSI. 
 The optimization problem of power consumption for BS is solved by 
 jointly optimizing the transmit precoding and the continuous reflect phase shifts based on the imperfect CSI and hardware impairments in an RIS-aided wireless communication system~\cite{9933780}.
 This paper~\cite{abeywickrama2020intelligent} formulates the optimization problem to minimize transmission power by jointly designing percoder and passive beamforming under the  phase-dependent amplitude model and perfect CSI, while satisfying individual signal-to-interference-plus-noise ratio constraints.
 This paper~\cite{saglam2022deep} presents a novel deep reinforcement learning DRL framework to maximize the sum downlink rate under imperfect CSI while considering hardware impairments and practical RIS phase-dependent amplitude model.

\textcolor{black}{
Obtaining precise CSI in such systems poses challenges,
which result in a prohibitive pilot overhead
~\cite{pan2022overview,wu2019intelligent,zheng2020intelligentofdm,9722893,9110587}.
Specifically, one aspect to consider is that with the growing number of IRS units, the dimensionality of channel estimation increases, leading to enhanced complexity.
Next, due to passive IRS channel, it is difficult to get segmented channel information. Additionally, the growing size of the IRS surface leads to near-field channel estimation issues. 
Moreover, hardware errors can negatively impact estimation accuracy, while beam offset can occur in IRS related channels. 
Therefore, these factors render the ideal CSI of cascaded channels unattainable in practical applications.}
\textcolor{black}{
	In addition, treating the estimated CSI as perfect through RIS in MIMO systems will result in inaccurate performance evaluation of the system.
 In the IRS-empowered systems, the significance of this issue is further amplified due to the need for estimation of additional IRS-related channels. 
 As a result, it becomes crucial to integrate content placement and transmission co-design that consider the presence of imperfect CSI. 
Initially, early research primarily focused on the imperfections in IRS-users channels while assuming perfect BS-IRS channels~\cite{pan2022overview,wu2021intelligent,9180053,9722893,9110587}. However, this approach necessitates separate estimation of BS-RIS-user channels,
which can be hardly implemented in practical.
Here, the work~\cite{9110587} introduces bound channel error models for worst-case
beamforming design by using independent RIS-related CSI. However, achieving this independent estimation is challenging. This approach requires separate estimation of IRS related channels, leading to implementation challenges and high complexity in solving beamforming problems.
Also, the authors did not consider cache optimization strategies.
Moreover, RIS-aided caching system optimizes content placement, active/passive beamforming under perfect CSI~\cite{9457078,li2022phase}.
} 

Motivated by the aforementioned observations, we consider caching 
content placement, active and passive beamforming co-design under 
imperfect CSI to reduce backhaul burden, enhance
 transmission robustness and beamforming gains. 
Main contributions are outlined as follows:
\begin{itemize}
		\item {
			Initially, a joint design of caching and communication under imperfect CSI is proposed in IRS-assisted MIMO networks for IoT. 
			Specifically, 		
			minimizing network costs includes both backhaul costs and transmission power, which are solved by optimizing active/passive beamforming and content placement under the imperfect CSI.
Several unit-modulus constraints and inequality constraints as well as the coupling of multiple variables make it difficult to solve the co-design problem directly. 
To handle this problem, we decouple this co-design problem (i.e. content placement, beamforming) by analyzing the structure of the optimization problem.
    Meanwhile, content placement problem can be addressed by applying the KKT optimality conditions.
	}
	\item { Then, worst-case robust beamforming design involves formulating minimizing transmit power problem. This is solved by joint optimizing passive phase vectors and the precoder under constant modulus constraints and worst-case QoS requirements. 
		However, this optimization task is challenging due to the strong coupling among variables, non-convex semi-infinite inequality constraints and constant modulus constraints. 
		In order to decouple these variables 
		in worst-case useful signal power inequality and interference-plus-noises (INs) power inequality, we employ techniques such as the S-procedure, Schur complement, and Sign-definiteness lemma. Therefore, these inequality constraints are linearly approximated as solvable equivalent constraints.
    }
	\item {
	 Additionally, we propose an efficient scheme that revolves around the iterative alternating optimization of the passive beamforming and precoder. 
	 To obtain suboptimal solutions, we introduce CCP
	 method and employ CVX techniques under the alternating optimization (AO) framework. By incorporating these techniques, we can achieve feasible solutions that are sufficiently close to the optimal ones.
 }
	\item {
		Finally,
		this work focuses on the collaborative design of content placement and hybrid beamforming under imperfect CSI. We propose an alternating optimization scheme to make a balance between power minimization and network cost minimization. We validate the proposed scheme and observe several key findings through simulations. Results demonstrate that the performance of our scheme surpasses that of uniform caching and random phase approaches, validating the effectiveness of the cooperative caching and transmission techniques. Further, the integration of IRS technology into caching to achieve power-saving and cost reducing benefits.
	}
\end{itemize}

The remainder of this manuscript is structured as follows:
In Section II, we consider a IRS enhanced edge caching system in an IoT networks, channel uncertainties and CSI error models, optimization problems.
In Section III, the content placement and joint beamforming co-design and the optimization scheme are investigated. 
Finally, the results and conclusions are presented in Section IV and Section V, respectively.

\textbf{Notations:} 
Scalars are denoted by italic letters. vectors and matrices are denoted by bold-face lower-case and upper-case letters, respectively.
${{\bf{A}}^{\mathrm{*}}}$, ${{\bf{A}}^{\mathrm{T}}}$ and ${{\bf{A}}^{\mathrm{H}}}$ denote the conjugate, transpose and conjugate transpose of matrix ${\bf{A}}$, respectively. 
$\mathbb{H}^{n}$ is $n \times n $ hermitian matrices.
$ \mathbb{H}_{+}^{n}$  is positive semidefinite matrices.
$ \mathbb{H}_{++}^{n}$ is positive definite matrices.
$\operatorname{diag}\{\cdot\}$ is diagonalization operation.
$\mathbb{E}[\cdot]$ is expectation operation. Independent and identically distributed is abbreviated as i.i.d..
$\mathcal{C} \mathcal{N}(\mu, \sigma^{2})$ is 
the circularly symmetric complex Gaussian distribution with mean ${\bf{\mu}}$ and variance $\sigma^{2}$.

   \vspace{-0.5cm}
\section{System Model}

We establish a system model and design an optimization problem for joint content placement and beamforming under imperfect CSI to test the robustness and reliability of
IRS-assisted caching system-oriented IoT networks.
\begin{figure}[t!]
	\begin{center}
		\includegraphics[width=3.2in]{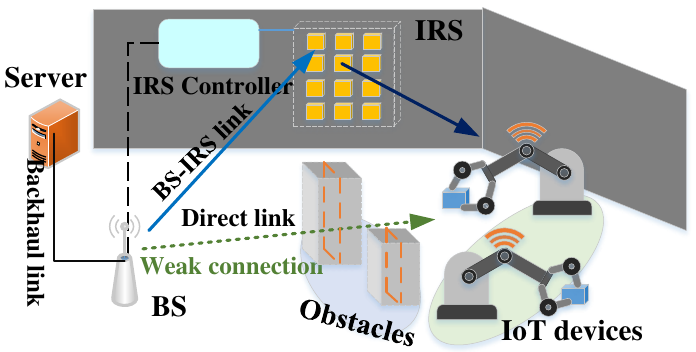}
		 \caption{IRS oriented edge caching system in an IoT network.}
		\label{fig_1}
	\end{center}
\end{figure}
\vspace{-0.5cm}
\subsection{System Description}
 In Fig. 1,
 the base station equipped with $N$ antennas concurrently serve $K$ single-antenna users (i.e., actuators, IoT devices).
  Meanwhile, base stations are connected to servers via high-capacity backhaul links and have access to databases containing files worth $F$.
Because of significant blockage resulting from large metallic machinery in manufacturing plants, the direct links between the BS and the devices experience considerable weakening and even loss of signal strength. An alternative route can be created through the installation of an IRS on either
the walls or ceiling to establish communication links~\cite{9599656}.
Additionally, we consider that the number of IRS phase shifters corresponds to $M$ reflecting elements. Furthermore, similar considerations are suitable for the uplink.
In addition, we have assumed that the quasi-static flat fading model applies to all channels~\cite{9180053,9933780}. 
In vertical industry applications like industrial IoT, manufacturing, and process automation, the mobility of devices is often limited~\cite{9599656}.
Specifically, devices in industrial settings usually move at low speeds or remain stationary, resulting in slow variations in the wireless channel over time.
Therefore, due to the limited mobility and slow channel variations, the fading experienced by the communication signals can be considered quasi-static, where the channel remains relatively constant over the duration of a communication block~\cite{9180053}. 
Additionally, IRS only performs first-order reflection.
 
\textbf{(1) Cache Model:}
The content $f$ can be accessed directly during the content fetching phase if it is cached in the BS's local storage. Otherwise, it must be fetched from BS over the backhaul
~\cite{9457078,li2022phase}. To ensure the data rate meets content delivery target rate $R_{k}^{0}$, total backhaul cost is modeled as  
$\sum_{f=1}^{F} \sum_{k=1}^{K}\left(1-c_{f}\right) b_{f} R_{k}^{0}$. 

In the content database,  a set of $F $ files  can be represented  $\mathcal{F}=\{1, \cdots, F\}$.
Then, each file has a standardized size of 1. 
We assume all files are sorted according to their popularity.
 Specifically, Zipf distribution is used as the distribution of popularity for each content file with a skewness factor of $\varepsilon$.
Next, $S_{0}$ is  local storage size, and $S_{0}$ cannot exceed $F$. Assuming BS adopts a probabilistic caching strategy. Further, $\mathbf{c}=\left\{c_{1}, \cdots, c_{f}, \cdots, c_{F}\right\} $ is content placement vector. Here,  $ c_{f} \in[0,1]$ denotes the probability that the $f$-th content file is cached.
Subject to this cache size limitation, $\sum_{f \in \mathcal{F}} c_{f} \leq S_{0}$ must be satisfied. Here, $b_{f}^{k}$ represents the probability that user $k$ requests file $f$. 
Note that assuming a unified distribution of user requests is adopted,
 i.e.,  $b_{f}^{1}=b_{f}^{2}=\cdots=b_{f}^{K} \triangleq b_{f} $.
Then, future work can also be extended to different user request distributions.
Furthermore, assuming that users request content files based on their popularity, the request probability follows Zipf distribution,
 characterized by a skewness factor $\varepsilon$, i.e.,  $b_{f}=\frac{f^{-\varepsilon}}{\sum_{i=1}^{F} i^{-\varepsilon}}$.
Then, a higher value of $\varepsilon$ indicates that more user requests are concentrated on a smaller number of popular files.

\textbf{(2) Communication Model:}
BS transmits Gaussian symbols $\mathbf{s} = \left[s_{1}, \ldots, s_{K}\right]^{\mathrm{T}} \in \mathbb{C}^{K \times 1}$ 
to receivers.
 Then, $\mathbb{E}\left[\mathbf{ss}^{\mathrm{H}}\right] = \mathbf{I}$ is covariance matrix. Additionly, the received baseband signal can be written as.
\begin{equation}
	\begin{aligned}\label{}
		y_{k}=\left(\mathbf{h}_{k}^{\mathrm{H}}+
		\mathbf{h}_{\mathrm{r}, k}^{\mathrm{H}} 
		\mathbf{\Theta} \mathbf{H}_{\mathrm{b,r}}
		\right) \mathbf{x}+n_{k}, \forall k \in \mathcal{K},
	\end{aligned}
\end{equation} 
where transmission signal $\mathbf{x} = \mathbf{W}\mathbf{s} = \sum_{k=1}^{K} \boldsymbol{w}_{k} s_{k}$. Here, the active beamforming is $\mathbf{W}$, $\mathbf{s}$ denotes the data symbols transmitted by the BS.
 Next, total transmission power limit is $P=\|\mathbf{W}\|_{F}^{2}$. In addition, this equation signifies that the transmitted signal is formed by the linear combination of the beamforming vectors weighted by the corresponding data symbols. Furthermore, 
the noise is $n_{k} \sim \mathcal{C N}\left(0, \sigma_{k}^{2}\right)$.

Next, 
 $\mathbf{\Theta}=\operatorname{diag}\left\{e^{j \theta_{1}}, \cdots, e^{j \theta_{m}}, \cdots, e^{j \theta_{M}}\right\}$  denote reflection matrix, where  $\theta_{m} \in[0,2 \pi) $ is the phase shift of the $m$-th reflecting element.  
Assume that IRS's phase shifts are calculated by transmitter and then transmitted to IRS controller via specific feedback channels~\cite{9180053,you2020channel,zhou2021joint,wang2020channel}.

Then, 
the channel to the user can be divided into two parts. Firstly, one part comes from the BS side (i.e., $\mathbf{h}_{k} \in \mathbb{C}^{N \times 1}$), and secondly, the remaining part is a cascaded channel formed by IRS incidence (i.e., $\mathbf{H}_{\mathrm{b,r}} \in \mathbb{C}^{M \times N}$) and reflection (i.e., $\mathbf{h}_{\mathrm{r}, k} \in \mathbb{C}^{M \times 1}$).

Then, the achievable rate is
\begin{equation}
\begin{aligned}\label{}
	\mathcal{R}_{k} =& B\log _{2}\left(1+\operatorname{SINR}_{k}\right)
\end{aligned}
\end{equation} 

Next, the SINR is represented as
\begin{equation}
	\begin{aligned}\label{}
 \operatorname{SINR}_{k}
=&\frac{\left|\left(\mathbf{e}^{\mathrm{H}} \mathbf{G}_{k}+\mathbf{h}_{k}^{\mathrm{H}}\right) \boldsymbol{w}_{k}\right|^{2}}{\left\|\left(\mathbf{h}_{k}^{\mathrm{H}}+\mathbf{e}^{\mathrm{H}} \mathbf{G}_{k}\right) \mathbf{W}_{-k}\right\|_{2}^{2}+\sigma_{k}^{2}}\\
=&\frac{\left|
	\left(\mathbf{h}_{k}^{\mathrm{H}}+\mathbf{e}^{\mathrm{H}} \mathbf{G}_{k}\right) \boldsymbol{w}_{k}\right|^{2}}
{\sum_{j \neq k}^{K}\left|
	\left(\mathbf{h}_{k}^{\mathrm{H}}+\mathbf{e}^{\mathrm{H}} \mathbf{G}_{k}\right) \boldsymbol{w}_{j}\right|^{2}+\sigma_{k}^{2}} 
\end{aligned}
\end{equation} 

Then, the interference-plus-noise power can be denoted as
\begin{equation}
	\begin{aligned}\label{}
IN_{k}= &{\sum_{j \neq k}^{K}\left|
	\left(\mathbf{h}_{k}^{\mathrm{H}}+\mathbf{e}^{\mathrm{H}} \mathbf{G}_{k}\right) \boldsymbol{w}_{j}\right|^{2}+\sigma_{k}^{2}}\\
	=&\left\|\left(\mathbf{h}_{k}^{\mathrm{H}}+\mathbf{e}^{\mathrm{H}} \mathbf{G}_{k}\right) \mathbf{W}_{-k}\right\|_{2}^{2}+\sigma_{k}^{2},
\end{aligned}
\end{equation} 
where $\mathbf{W}_{-k}=\left[\boldsymbol{w}_{1}, \ldots, \boldsymbol{w}_{k-1}, \boldsymbol{w}_{k+1}, \cdots, \boldsymbol{w}_{K}\right]$. $B$ is the system bandwidth.
Here, the cascaded channel through the IRS can be denoted as
\begin{equation}
	\begin{aligned}\label{}
 \mathbf{G}_{k}=\operatorname{diag}\left(\mathbf{h}_{\mathrm{r}, k}^{\mathrm{H}}\right) \mathbf{H}_{\mathrm{b,r}}
\end{aligned}
\end{equation} 

\subsection{Channel Uncertainties and Error Models}
  
  Obtaining the cascaded channels through IRS is more difficult compared to acquiring the channel knowledge of the direct channel.
   Here, assuming that direct channels are perfectly known while the IRS-related channels are imperfect.
  To model channel uncertainty, we employ a norm bounded additive uncertainty set~\cite{9180053,chi2017convex}. 
  Therefore, we can obtain  
  \begin{equation}
  \begin{aligned}\label{}
  \mathbf{G}_{k}=\widehat{\mathbf{G}}_{k}+\triangle \mathbf{G}_{k}, \forall k \in \mathcal{K},
  \end{aligned}
  \end{equation} 
where $\widehat{\mathbf{G}}_{k}$ denotes the transmitter's estimation of the channel for user $k$, while $\triangle \mathbf{G}_{k}$ represents associated estimation error.
Prior studies on transceiver design have often operated under the ideal assumption of perfect CSI~\cite{wu2020towards,wu2019intelligent}. However, this assumption is not feasible in practice, due to the propagation of channel estimation errors within IRS-related channels.
 Assuming that it is contained within a ball of the given radius $\xi_{\mathrm{g}, k}$, i.e, $\left\|\triangle \mathbf{G}_{k}\right\|_{F} \leq \xi_{\mathrm{g}, k}$.
 The expression $\left\|\triangle \mathbf{G}_{k}\right\|_{F} \leq \xi_{\mathrm{g}, k}$ specifies a norm constraint on the uncertainty vector $\left\|\triangle \mathbf{G}_{k}\right\|_{F}$ for user $k$, $\xi_{\mathrm{g}, k}$ represents the radius of uncertainty region known at BS.
For worst-case robust approach, CSI errors are assumed to reside within a bounded set. The objective is to develop precoder to be resilient against worst-case QoS under specified CSI error model~\cite{chi2017convex}.
Then, this model is particularly applicable in systems where CSI is quantized at the receivers and transmitted back to transmitter in communication, as described in~\cite{9180053,chi2017convex,gharavol2013the}. 
More specifically, in frequency division duplexing systems, the estimation of the channel 
 is crucial, and it is achieved by utilizing a feedback channel from the receiver.
However, due to the limited capacity of feedback channel, channel response needs to be quantized, which introduces CSI errors.
In particular, when an almost uniform quantizer is employed, it is possible to approximate quantization cells within the interior of quantization region as balls with a size of $\xi_{\mathrm{g}, k}$~\cite{9180053,chi2017convex,gharavol2013the,4407774,1561600}.

\subsection{Problem Formulation}

The aim is to minimize network cost 
by optimizing content placement vector, precoder and reflection beamforming. 
The main concept is to separate the optimization of content placement from joint beamforming design, and then separate the precoder from passive beamforming in order to iteratively update each other until convergence.
Then, the optimization task is conducted while adhering to constant modulus constraints as well as ensuring worst-case QoS.
The QoS entail guaranteeing that the achievable rate of each user in the IoT networks remains above a threshold for all conceivable channel error realizations~\cite{9180053,chi2017convex,4407774,9110587}.
Moreover, let $\mu_{k} \triangleq \left\{\forall\left\|\triangle \mathbf{G}_{k}\right\|_{F} \leq \xi_{\mathrm{g}, k}\right\}$, $\mathcal{M}=\{1,2, \ldots, M\}$. 
Then, we can obtain 
\begin{subequations}
	\begin{align}
	\mathcal{P}_{1}:
	\min _{\left\{c_{f}\right\}, \mathbf{W}, \mathbf{e}} &
	 \sum_{f=1}^{F} \sum_{k=1}^{K}\left(1-c_{f}\right) b_{f} R_{k}^{0}+ \eta \sum_{k=1}^{K}\left\|\boldsymbol{w}_{k}\right\|^{2}\\
	\text { s.t. } & B\log _{2}\left(1+\operatorname{SINR}_{k}\right)
	 \geq \gamma_{k}, \mu_{k}, \forall k \in \mathcal{K} \\
	& \left|e_{m}\right|^{2}=1, \forall m \in \mathcal{M},\\
	&	c_{f} \in[0,1], \quad \forall f \in \mathcal{F} \\
	&	\sum_{f=1}^{F} c_{f} \leq S_{0},
	\end{align}
\end{subequations}

where 
target rate is $\gamma_{k}$. Constraints (7b) define worst-case QoS requirements. Then, (7c) is the unit-modulus.
Although the objective function in (7a) has convexity, solving this problem is difficult due to non convex constraints and coupling between precoder and passive beamforming vectors.
It is worth noting that there is no standard method available for optimal solutions to $	\mathcal{P}_{1}$~\cite{chi2017convex}. Constraint (7e) considers the local
storage limit at the transmitter. (7d) indicates the
probability that the $f$-th content file is cached.

\section{Alternating Optimization Framework}

Alternating optimization approach is designed to address network cost minimization problem. 

\subsection{Content Placement } 
When fixing precoder and reflecting beamforming, $\mathcal{P}_{1}$ is transformed into

\begin{subequations}
	\begin{align}
		\mathcal{P}_{2}:\min _{\left\{c_{f}\right\}} & \sum_{f=1}^{F}\left(1-c_{f}\right) b_{f} \\
		\text { s.t. }& c_{f} \in[0,1], \quad \forall f \in \mathcal{F} \\
		&\sum_{f=1}^{F} c_{f} \leq S_{0} 
	\end{align}
\end{subequations}

$\mathcal{P}_{2}$ aims to find the optimal caching strategy. 
We adopt the karush-kuhn-tucker (KKT) optimality condition to solve the content placement vectors~\cite{chi2017convex}.

\subsection{Active and Passive Beamforming}  
Further, the worst-case joint beamforming problem is written as
\begin{subequations}
	\begin{align}
	\text {(P1')}:
	\min _{\mathbf{W}, \mathbf{e}} & \sum_{k=1}^{K}\left\|\boldsymbol{w}_{k}\right\|^{2}\\
	\text { s.t. } & \mathcal{R}_{k} \geq \gamma_{k}, \mu_{k}, \forall k \in \mathcal{K} \\
	& \left|e_{m}\right|^{2}=1, \forall m \in \mathcal{M}
	\end{align}
\end{subequations}

Additionally, unit-modulus constraint is (9c). 
These constraints enforce that the magnitude of the reflection coefficients remains unity.
Then, constraints (9b) in the problem formulation represent the worst-case QoS requirements. These consist of two inequality constraints. The purpose of these constraints is to ensure that the system meets the desired QoS requirements for all users.
\begin{equation}
\begin{aligned}
\left|\left(\mathbf{h}_{k}^{\mathrm{H}}+\mathbf{e}^{\mathrm{H}} \mathbf{G}_{k}\right) \boldsymbol{w}_{k}\right|^{2} \geq IN_{k} \left(2^{\gamma_{k}/B}-1\right), \mu_{k}, \forall k \in \mathcal{K},
\end{aligned}
\end{equation}
\begin{equation}
\begin{aligned}
\left\|\left(\mathbf{h}_{k}^{\mathrm{H}}+\mathbf{e}^{\mathrm{H}} \mathbf{G}_{k}\right) \mathbf{W}_{-k}\right\|_{2}^{2}+\sigma_{k}^{2} \leq IN_{k}, \mu_{k}, \forall k \in \mathcal{K},
\end{aligned}
\end{equation}
where we view the INs power $\mathbf {IN}=\left[IN_{1}, \ldots, IN_{K}\right]^{\mathrm{T}}$ as auxiliary variables.
	Then, from the properties of logarithmic functions, 
	$\mathcal{R}_{k} \geq \gamma_{k}$ can be concluded that
\begin{equation}
\begin{aligned}
B \log _{2}\left(1+\frac{1}{IN_{k}}\left|\left(\mathbf{h}_{k}^{\mathrm{H}}+\mathbf{e}^{\mathrm{H}} \mathbf{G}_{k}\right) \boldsymbol{w}_{k}\right|^{2}\right)\geq \gamma_{k}\\
\left|\left(\mathbf{h}_{k}^{\mathrm{H}}+\mathbf{e}^{\mathrm{H}} \mathbf{G}_{k}\right) \boldsymbol{w}_{k}\right|^{2} \geq IN_{k}\left(2^{\gamma_{k}/B}-1\right).
\end{aligned}
\end{equation}
 
Then, 
we obtain constraint (10) by following the steps below.
Firstly, we approximate the non-convex parts of (10) to make them amenable to optimization. 
Secondly,
we solve semi-infinite inequalities by  S-procedure~\cite{chi2017convex,boyd2004convex}.
By employing these approach, we effectively resolve the non-convexity of the problem.

First-order Taylor inequality~\cite{9180053}:
If $c$ is a complex scalar variable, then there is an inequality.
 \begin{align}\label{gamma first-order Taylor inequality}
 |c|^{2} \geq 2 \operatorname{Re}\left\{c^{*,(n)} c\right\}-c^{*,(n)} c^{(n)}
 \end{align} 

In addition, there is the following expression through using the other inequalities mentioned above.
 \begin{align}\label{gamma new Taylor inequality}
 	\begin{array}{l}
 		\left|\left(\mathbf{h}_{k}^{\mathrm{H}}+\mathbf{e}^{\mathrm{H}} \mathbf{G}_{k}\right) \boldsymbol{w}_{k}\right|^{2} \\
 		\geq 2 \operatorname{Re}\left\{\left(\mathbf{h}_{k}^{\mathrm{H}}+\mathbf{e}^{(n), \mathrm{H}} \mathbf{G}_{k}\right) \boldsymbol{w}_{k}^{(n)} \boldsymbol{w}_{k}^{\mathrm{H}}\left(\mathbf{h}_{k}+\mathbf{G}_{k}^{\mathrm{H}} \mathbf{e}\right)\right\} \\
 		-\left(\mathbf{h}_{k}^{\mathrm{H}}+\mathbf{e}^{(n), \mathrm{H}} \mathbf{G}_{k}\right) \boldsymbol{w}_{k}^{(n)} \boldsymbol{w}_{k}^{(n), \mathrm{H}}\left(\mathbf{h}_{k}+\mathbf{G}_{k}^{\mathrm{H}} \mathbf{e}^{(n)}\right), 
 	\end{array}
 \end{align}
By substituting $\mathbf{G}_{k}=\widehat{\mathbf{G}}_{k}+\Delta \mathbf{G}_{k}$ into (14)
and simplifying it through some special mathematical transformations,  i.e., 
$\operatorname{Tr}\left(\mathbf{A B C D}\right)=\left(\operatorname{vec}^{\mathrm{T}}(\mathbf{D})\right)^{\mathrm{T}}\left(\mathbf{C}^{\mathrm{T}} \otimes \mathbf{A}\right) \operatorname{vec}(\mathbf{B})$, $\operatorname{Tr}\left(\mathbf{A}^{\mathrm{H}} \mathbf{B}\right)=\operatorname{vec}^{\mathrm{H}}(\mathbf{A}) \operatorname{vec}\left(\mathbf{B}\right)$~\cite{9180053,9110587,zhang2017matrix}.
Next, we considered the derivation process and obtained the first term (14).
\begin{align*} 
		&\left(\mathbf{h}_{k}^{\mathrm{H}}+\mathbf{e}^{(n), \mathrm{H}} (\widehat{\mathbf{G}}_{k}+\triangle \mathbf{G}_{k})\right) \boldsymbol{w}_{k}^{(n)} \boldsymbol{w}_{k}^{\mathrm{H}}	
		\left(\mathbf{h}_{k}+
		(\widehat{\mathbf{G}}_{k}+\triangle \mathbf{G}_{k})^{\mathrm{H}} \mathbf{e}\right) 
		\\
		&=\mathbf{h}_{k}^{\mathrm{H}} \boldsymbol{w}_{k}^{(n)} \boldsymbol{w}_{k}^{\mathrm{H}} (\mathbf{h}_{k}+
		\widehat{\mathbf{G}}_{k}^{\mathrm{H}} \mathbf{e}
		+ \triangle \mathbf{G}_{k}^{\mathrm{H}} \mathbf{e})
		+\mathbf{e}^{(n), \mathrm{H}} \widehat{\mathbf{G}}_{k} 
		\boldsymbol{w}_{k}^{(n)} \boldsymbol{w}_{k}^{\mathrm{H}} (\mathbf{h}_{k}+
		\widehat{\mathbf{G}}_{k}^{\mathrm{H}} \mathbf{e}
		+ \triangle \mathbf{G}_{k}^{\mathrm{H}} \mathbf{e})
		\\
		& +\mathbf{e}^{(n), \mathrm{H}} \triangle \mathbf{G}_{k}^{\mathrm{H}} 
		\boldsymbol{w}_{k}^{(n)} \boldsymbol{w}_{k}^{\mathrm{H}} (\mathbf{h}_{k}+
		\widehat{\mathbf{G}}_{k}^{\mathrm{H}} \mathbf{e}
		+ \triangle \mathbf{G}_{k}^{\mathrm{H}} \mathbf{e})
	\\
		&=\mathbf{h}_{k}^{\mathrm{H}} \boldsymbol{w}_{k}^{(n)} \boldsymbol{w}_{k}^{\mathrm{H}} \mathbf{h}_{k}
		+\mathbf{h}_{k}^{\mathrm{H}} \boldsymbol{w}_{k}^{(n)} \boldsymbol{w}_{k}^{\mathrm{H}} (\widehat{\mathbf{G}}_{k}^{\mathrm{H}} \mathbf{e})
		+\mathbf{h}_{k}^{\mathrm{H}} \boldsymbol{w}_{k}^{(n)} \boldsymbol{w}_{k}^{\mathrm{H}} (\triangle \mathbf{G}_{k}^{\mathrm{H}} \mathbf{e})
		\\
		& +\mathbf{e}^{(n), \mathrm{H}} \widehat{\mathbf{G}}_{k} 
		\boldsymbol{w}_{k}^{(n)} \boldsymbol{w}_{k}^{\mathrm{H}} \mathbf{h}_{k}
		+\mathbf{e}^{(n), \mathrm{H}} \widehat{\mathbf{G}}_{k} 
		\boldsymbol{w}_{k}^{(n)} \boldsymbol{w}_{k}^{\mathrm{H}}
		(\widehat{\mathbf{G}}_{k}^{\mathrm{H}} \mathbf{e})
	+\mathbf{e}^{(n), \mathrm{H}} \widehat{\mathbf{G}}_{k} 
		\boldsymbol{w}_{k}^{(n)} \boldsymbol{w}_{k}^{\mathrm{H}}
		(\triangle \mathbf{G}_{k}^{\mathrm{H}} \mathbf{e}) 	
		\\
		& +\mathbf{e}^{(n), \mathrm{H}} \triangle \mathbf{G}_{k}^{\mathrm{H}} 
		\boldsymbol{w}_{k}^{(n)} \boldsymbol{w}_{k}^{\mathrm{H}} (\mathbf{h}_{k})+
		\mathbf{e}^{(n), \mathrm{H}} \triangle \mathbf{G}_{k}^{\mathrm{H}} 
		\boldsymbol{w}_{k}^{(n)} \boldsymbol{w}_{k}^{\mathrm{H}}
		(\widehat{\mathbf{G}}_{k}^{\mathrm{H}} \mathbf{e})
		 +\mathbf{e}^{(n), \mathrm{H}} \triangle \mathbf{G}_{k}^{\mathrm{H}} 
		\boldsymbol{w}_{k}^{(n)} \boldsymbol{w}_{k}^{\mathrm{H}}
		(\triangle \mathbf{G}_{k}^{\mathrm{H}} \mathbf{e})
\end{align*}

 Next, we get the second term.  
\begin{align*}
&\left(\mathbf{h}_{k}^{\mathrm{H}}+\mathbf{e}^{(n), \mathrm{H}} \mathbf{G}_{k}\right) \boldsymbol{w}_{k}^{(n)} \boldsymbol{w}_{k}^{(n), \mathrm{H}}\left(\mathbf{h}_{k}+\mathbf{G}_{k}^{\mathrm{H}} \mathbf{e}^{(n)}\right)
\\
&=\left(\mathbf{h}_{k}^{\mathrm{H}}+\mathbf{e}^{(n), \mathrm{H}} (\widehat{\mathbf{G}}_{k}+\triangle \mathbf{G}_{k})\right) \boldsymbol{w}_{k}^{(n)} \boldsymbol{w}_{k}^{(n), \mathrm{H}}
\left(\mathbf{h}_{k}+(\widehat{\mathbf{G}}_{k}+\triangle \mathbf{G}_{k})^{\mathrm{H}} \mathbf{e}^{(n)}\right)
\\
&=\left(\mathbf{h}_{k}^{\mathrm{H}}+\mathbf{e}^{(n), \mathrm{H}} \widehat{\mathbf{G}}_{k}
+\mathbf{e}^{(n), \mathrm{H}} \triangle \mathbf{G}_{k}\right)
\boldsymbol{w}_{k}^{(n)} \boldsymbol{w}_{k}^{(n), \mathrm{H}}
\left(\mathbf{h}_{k}+\widehat{\mathbf{G}}_{k}^{\mathrm{H}} \mathbf{e}^{(n)}+
\triangle \mathbf{G}_{k}^{\mathrm{H}} \mathbf{e}^{(n)}\right)
\\
&=\mathbf{h}_{k}^{\mathrm{H}} \boldsymbol{w}_{k}^{(n)} \boldsymbol{w}_{k}^{(n), \mathrm{H}}
(\mathbf{h}_{k}+\widehat{\mathbf{G}}_{k}^{\mathrm{H}} \mathbf{e}^{(n)}+
\triangle \mathbf{G}_{k}^{\mathrm{H}} \mathbf{e}^{(n)})
+\mathbf{e}^{(n), \mathrm{H}} \widehat{\mathbf{G}}_{k}
\boldsymbol{w}_{k}^{(n)} \boldsymbol{w}_{k}^{(n), \mathrm{H}}
(\mathbf{h}_{k}+\widehat{\mathbf{G}}_{k}^{\mathrm{H}} \mathbf{e}^{(n)}+
\triangle \mathbf{G}_{k}^{\mathrm{H}} \mathbf{e}^{(n)})\\
&
+\mathbf{e}^{(n), \mathrm{H}} \triangle \mathbf{G}_{k}  \boldsymbol{w}_{k}^{(n)} \boldsymbol{w}_{k}^{(n), \mathrm{H}}
(\mathbf{h}_{k}+\widehat{\mathbf{G}}_{k}^{\mathrm{H}} \mathbf{e}^{(n)}+
\triangle \mathbf{G}_{k}^{\mathrm{H}} \mathbf{e}^{(n)})
\\
&=\mathbf{h}_{k}^{\mathrm{H}} \boldsymbol{w}_{k}^{(n)} \boldsymbol{w}_{k}^{(n), \mathrm{H}}
(\mathbf{h}_{k})
+\mathbf{h}_{k}^{\mathrm{H}} \boldsymbol{w}_{k}^{(n)} \boldsymbol{w}_{k}^{(n), \mathrm{H}}(\widehat{\mathbf{G}}_{k}^{\mathrm{H}} \mathbf{e}^{(n)})
+\mathbf{h}_{k}^{\mathrm{H}} \boldsymbol{w}_{k}^{(n)} \boldsymbol{w}_{k}^{(n), \mathrm{H}}
(\triangle \mathbf{G}_{k}^{\mathrm{H}} \mathbf{e}^{(n)})
\\
&+\mathbf{e}^{(n), \mathrm{H}} \widehat{\mathbf{G}}_{k}
\boldsymbol{w}_{k}^{(n)} \boldsymbol{w}_{k}^{(n), \mathrm{H}}
(\mathbf{h}_{k})
+\mathbf{e}^{(n), \mathrm{H}} \widehat{\mathbf{G}}_{k}
\boldsymbol{w}_{k}^{(n)} \boldsymbol{w}_{k}^{(n), \mathrm{H}}(\widehat{\mathbf{G}}_{k}^{\mathrm{H}} \mathbf{e}^{(n)})
+\mathbf{e}^{(n), \mathrm{H}} \widehat{\mathbf{G}}_{k}
\boldsymbol{w}_{k}^{(n)} \boldsymbol{w}_{k}^{(n), \mathrm{H}}
(\triangle \mathbf{G}_{k}^{\mathrm{H}} \mathbf{e}^{(n)})
\\
&+\mathbf{e}^{(n), \mathrm{H}} \triangle \mathbf{G}_{k}  \boldsymbol{w}_{k}^{(n)} \boldsymbol{w}_{k}^{(n), \mathrm{H}}
(\mathbf{h}_{k})
+\mathbf{e}^{(n), \mathrm{H}} \triangle \mathbf{G}_{k}  \boldsymbol{w}_{k}^{(n)} \boldsymbol{w}_{k}^{(n), \mathrm{H}}
(\widehat{\mathbf{G}}_{k}^{\mathrm{H}} \mathbf{e}^{(n)})
+\mathbf{e}^{(n), \mathrm{H}} \triangle \mathbf{G}_{k}  \boldsymbol{w}_{k}^{(n)} \boldsymbol{w}_{k}^{(n), \mathrm{H}}
(\triangle \mathbf{G}_{k}^{\mathrm{H}} \mathbf{e}^{(n)})
\end{align*}

Then, by substituting the above expansion into the (14), 
$\mid[\mathbf{h}_{k}^{\mathrm{H}}+\mathbf{e}^{\mathrm{H}}
(\widehat{\mathbf{G}}_{k}+
\triangle \mathbf{G}_{k})]\left.\boldsymbol{w}_{k}\right|^{2}$
 is linearly approximated by its lower bound at  $(\boldsymbol{w}_{k}^{(n)}, \mathbf{e}^{(n)})$:
\begin{equation}
\begin{aligned}
\operatorname{vec}^{\mathrm{T}}\left(\triangle \mathbf{G}_{k}\right) \mathbf{J}_{k} \operatorname{vec}\left(\triangle \mathbf{G}_{k}^{*}\right)+2 \operatorname{Re}\left\{\mathbf{j}_{k}^{\mathrm{T}} \operatorname{vec}\left(\triangle \mathbf{G}_{k}^{*}\right)\right\}+j_{k},
\end{aligned}
\end{equation}
where 
 \begin{equation}
\begin{aligned}
\mathbf{J}_{k}&= \boldsymbol{w}_{k} \boldsymbol{w}_{k}^{(n), \mathrm{H}} \otimes \mathbf{e}^{*} \mathbf{e}^{(n), \mathrm{T}}+\boldsymbol{w}_{k}^{(n)} \boldsymbol{w}_{k}^{\mathrm{H}} \otimes \mathbf{e}^{(n), *} \mathbf{e}^{\mathrm{T}} \\
& -\left(\boldsymbol{w}_{k}^{(n)} \boldsymbol{w}_{k}^{(n), \mathrm{H}} \otimes \mathbf{e}^{(n), *} \mathbf{e}^{(n), \mathrm{T}}\right),\\
\mathbf{j}_{k}&=  \operatorname{vec}\left(\mathbf{e}\left(\mathbf{h}_{k}^{\mathrm{H}}+\mathbf{e}^{(n), \mathrm{H}} \widehat{\mathbf{G}}_{k}\right) \boldsymbol{w}_{k}^{(n)} \boldsymbol{w}_{k}^{\mathrm{H}}\right) \\
& +\operatorname{vec}\left(\mathbf{e}^{(n)}\left(\mathbf{h}_{k}^{\mathrm{H}}+\mathbf{e}^{\mathrm{H}} \widehat{\mathbf{G}}_{k}\right) \boldsymbol{w}_{k} \boldsymbol{w}_{k}^{(n), \mathrm{H}}\right) \\
& -\operatorname{vec}\left(\mathbf{e}^{(n)}\left(\mathbf{h}_{k}^{\mathrm{H}}+\mathbf{e}^{(n), \mathrm{H}} \widehat{\mathbf{G}}_{k}\right) \boldsymbol{w}_{k}^{(n)} \boldsymbol{w}_{k}^{(n), \mathrm{H}}\right),\\
j_{k}&= 2 \operatorname{Re}\left\{\left(\mathbf{h}_{k}^{\mathrm{H}}+\mathbf{e}^{(n), \mathrm{H}} \widehat{\mathbf{G}}_{k}\right) \boldsymbol{w}_{k}^{(n)} \boldsymbol{w}_{k}^{\mathrm{H}}\left(\mathbf{h}_{k}+\widehat{\mathbf{G}}_{k}^{\mathrm{H}} \mathbf{e}\right)\right\} \\
& -\left(\mathbf{h}_{k}^{\mathrm{H}}+\mathbf{e}^{(n), \mathrm{H}} \widehat{\mathbf{G}}_{k}\right) \boldsymbol{w}_{k}^{(n)} \boldsymbol{w}_{k}^{(n), \mathrm{H}}\left(\mathbf{h}_{k}+\widehat{\mathbf{G}}_{k}^{\mathrm{H}} \mathbf{e}^{(n)}\right).
\end{aligned}
\end{equation}

Then, we can obtain the constraints
\begin{equation}
\begin{aligned}
\operatorname{vec}^{\mathrm{T}}\left(\triangle \mathbf{G}_{k}\right) \mathbf{J}_{k} \operatorname{vec}\left(\Delta \mathbf{G}_{k}^{*}\right)+2 \operatorname{Re}\left\{\mathbf{j}_{k}^{\mathrm{T}} \operatorname{vec}\left(\triangle \mathbf{G}_{k}^{*}\right)\right\}+j_{k} \\
\geq IN_{k}\left(2^{\gamma _{k}/B}-1\right), \mu _{k}, \forall k \in \mathcal{K}
\end{aligned}
\end{equation}
 
 Currently, S-procedure is extensively employed in robust beamforming for MIMO  communications~\cite{chi2017convex}. 
 Lemma 1. (S-procedure~\cite{chi2017convex,boyd2004convex,boyd1994linear}) :
Let $ F_{0}, \ldots, F_{p} $ be quadratic functions of the variable $ \Psi \in \mathbf{R}^{n} $
\begin{equation}
\begin{aligned}
F_{i}(\Psi) \triangleq \Psi^{\mathrm{H}} \mathbf{U}_{i} \Psi+2 
\mathbf{u}_{i}^{\mathrm{H}} \Psi+v_{i}, \quad i=0, \ldots, p,
\end{aligned}
\end{equation}
where $\mathbf{U}_{i}=\mathbf{U}_{i}^{\mathrm{H}} $.
Then, the condition on $F_{0}, \ldots, F_{p} $:
\begin{equation}
\begin{aligned}
F_{0}(\Psi) \geq 0 \text { for } \Psi \text { such that } F_{i}(\Psi) \geq 0, \quad i=1, \ldots, p .
\end{aligned}
\end{equation}

There exist $ \tau_{1} \geq 0, \ldots, \tau_{p} \geq 0 $ such that
\begin{equation}
\begin{aligned}
\text { for every } \Psi \text {,}
\quad F_{0}(\Psi)-\sum_{i=1}^{p} \tau_{i} F_{i}(\Psi) \geq 0,
\end{aligned}
\end{equation}
then (20) holds. 
Note that (20) can be written as
\begin{equation}
\begin{aligned}
\left[\begin{array}{cc}
\mathbf{U}_{0} & 
\mathbf{u}_{0} \\
\mathbf{u}_{0}^{\mathrm{H}} & v_{0}
\end{array}\right]-\sum_{i=1}^{p} \tau_{i}\left[\begin{array}{cc}
\mathbf{U}_{i} & 
\mathbf{u}_{i} \\
\mathbf{u}_{i}^{\mathrm{H}} & v_{i}
\end{array}\right] \succeq \mathbf{0}
\end{aligned}
\end{equation}

Remark:
the (20) and (21) are equivalent under this condition (i.e., the functions  $F_{i}$  are affine).

By setting the parameters in Lemma 1, constraint (17) can be formulated as follows:
\begin{equation}
\begin{aligned}
\mathbf{U}_{0}=\mathbf{J}_{k}, \mathbf{u}_{0}=\mathbf{j}_{k},  v_{0}=j_{k}-IN_{k}\left(2^{\gamma_{k}/B}-1\right), \\
\quad \mathbf{U}_{1}=-\mathbf{I},
\mathbf{u}_{1}=\mathbf{0}, v_{1}=\xi_{k}^{2}, \mathbf{x}=\operatorname{vec}\left(\triangle \mathbf{G}_{k}^{*}\right).
\end{aligned}
\end{equation} 

Next, (17) is transformed into the following equivalent linear matrix inequality (LMI) as
\begin{subequations}
	\begin{align}
	\left[\begin{array}{cc}
	\mathbf{U}_{0}-\tau_{\mathrm{g}, k} \mathbf{U}_{1} & 
	\mathbf{u}_{0} -\tau_{\mathrm{g}, k} \mathbf{u}_{1} \\
	\mathbf{u}_{0}^{\mathrm{H}}-\tau_{\mathrm{g}, k} \mathbf{u}_{1}^{\mathrm{H}} & v_{0}-\tau_{\mathrm{g}, k} v_{1}
	\end{array}\right] \succeq \mathbf{0},\\
\left[\begin{array}{cc}
\tau_{\mathrm{g}, k} \mathbf{I}_{M N}+\mathbf{J}_{k} & \mathbf{j}_{k} \\
\mathbf{j}_{k}^{\mathrm{H}} & C_{k}
\end{array}\right] \succeq \mathbf{0}, 
\end{align}
\end{subequations}
where slack variables are  $\boldsymbol{\tau}_{\mathrm{g}}=\left[\tau_{\mathrm{g}, 1}, \ldots, \tau_{\mathrm{g}, K}\right]^{\mathrm{T}} \geq 0$ and  $C_{k}=j_{k}-IN_{k}\left(2^{\gamma_{k}/B}-1\right)-\tau_{\mathrm{g}, k} \xi_{k}^{2}.$
%
We introduce Schur complement to transform matrix inequalities~\cite{chi2017convex}.

 Lemma 2. (Schur complement~\cite{chi2017convex}) :
Suppose that  $\mathbf{C} \in \mathbb{H}_{++}^{m}$, $\mathbf{A} \in \mathbb{H}^{n}$, and  $\mathbf{B} \in \mathbb{C}^{n \times m}$. Then\begin{equation}
\begin{aligned}
\mathbf{S} \triangleq\left[\begin{array}{cc}
\mathbf{A} & \mathbf{B} \\
\mathbf{B}^{\mathrm{H}} & \mathbf{C}
	\end{array}\right] \succeq \mathbf{0},
	\end{aligned}
	\end{equation}
	where $\mathbf{S}_{\mathbf{C}} \triangleq \mathbf{A}-\mathbf{B C}^{-1} \mathbf{B}^{\mathrm{H}}\succeq \mathbf{0}$. 
 In order to effectively solve the robust beamforming problem, we use Schur complement to equivalently rewrite the INs power inequality 
 as a matrix inequality (24),
 as follows:
\begin{subequations}
	\begin{align}
	IN_{k}- \sigma_{k}^{2} -\left\|\left(\mathbf{h}_{k}^{\mathrm{H}}+\mathbf{e}^{\mathrm{H}} \mathbf{G}_{k}\right) \mathbf{W}_{-k}\right\|_{2}^{2} \ge 0, \forall k \in \mathcal{K},\\
	\left[\begin{array}{cc}
	IN_{k}-\sigma_{k}^{2} & \mathbf{T}_{k}^{\mathrm{H}} \\
	\mathbf{T}_{k} & \mathbf{I}
	\end{array}\right] \succeq \mathbf{0},
	\end{align}
\end{subequations}
where  $\mathbf{T}_{k}=\left(\left(\mathbf{h}_{k}^{\mathrm{H}}+\mathbf{e}^{\mathrm{H}} \mathbf{G}_{k}\right) \mathbf{W}_{-k}\right)^{\mathrm{H}}$.
Substituting $\mathbf{G}_{k}=\widehat{\mathbf{G}}_{k}+ \triangle \mathbf{G}_{k}$ into (25b), we obtain 
	\begin{equation}
	\begin{aligned}
	\begin{array}{l}
	{\left[\begin{array}{cc}
		IN_{k}- \sigma_{k}^{2} & \widehat{\mathbf{T}}_{k}^{\mathrm{H}} \\
		\widehat{\mathbf{T}}_{k} & \mathbf{I}
		\end{array}\right] \succeq-\left[\begin{array}{c}
		\mathbf{0} \\
		\mathbf{W}_{-k}^{\mathrm{H}}
		\end{array}\right] \triangle \mathbf{G}_{k}^{\mathrm{H}}\left[\begin{array}{ll}
		\mathbf{e} & \mathbf{0}
		\end{array}\right]} \\
	-\left[\begin{array}{c}
	\mathbf{e}^{\mathrm{H}} \\
	\mathbf{0}
	\end{array}\right] \triangle \mathbf{G}_{k}\left[\begin{array}{cc}
	\mathbf{0} & \mathbf{W}_{-k}
	\end{array}\right], \\
	\end{array}
	\end{aligned}
	\end{equation}
	where  $\widehat{\mathbf{T}}_{k}=((\mathbf{h}_{k}^{\mathrm{H}}+\mathbf{e}^{\mathrm{H}} \widehat{\mathbf{G}}_{k}) \mathbf{W}_{-k})^{\mathrm{H}}$.
	Next, Sign-definiteness lemma is adopted to simplify (26).
	
	Lemma 3. (Sign-definiteness lemma~\cite{gharavol2013the,eldar2005robust}) : Let  matrices $\mathbf{P}$, $\mathbf{Q}$, $\mathbf{A}$ ($\mathbf{A}=\mathbf{A}^{*}$).
	\begin{equation}
	\begin{aligned}
	\mathbf{A} \succeq \mathbf{P}^{*} \mathbf{X Q}+\mathbf{Q}^{*} \mathbf{X}^{*} \mathbf{P}, \quad \forall \mathbf{X}:\|\mathbf{X}\| \leq \rho
	\end{aligned}
	\end{equation}
	
	if and only if there exists a $\lambda \geq 0$ such that
	\begin{equation}
	\begin{aligned}
	\left[\begin{array}{cc}
	\mathbf{A}-\lambda \mathbf{Q}^{*} \mathbf{Q} & -\rho \mathbf{P}^{*} \\
	-\rho \mathbf{P} & \lambda \mathbf{I}
	\end{array}\right] \succeq 0 .
	\end{aligned}
	\end{equation}

	The corresponding proof can be found in the literature~\cite{chi2017convex,boyd2004convex,gharavol2013the,khlebnikov2008petersens,eldar2005robust,boyd1994linear}.
	Referring to Sign-definiteness lemma, we choose the following parameters:
	\begin{equation}
	\begin{aligned}
	\begin{array}{l}
	\mathbf{A}=\left[\begin{array}{cc}
	IN_{k}-\sigma_{k}^{2} & \widehat{\mathbf{T}}_{k}^{\mathrm{H}} \\
	\widehat{\mathbf{T}}_{k} & \mathbf{I}
	\end{array}\right], 
	\mathbf{P}=-\left[\begin{array}{cc}
	\mathbf{0} & \mathbf{F}_{-k}
	\end{array}\right], \\
	\mathbf{Q}=\left[\begin{array}{ll}
	\mathbf{e} & \mathbf{0}
	\end{array}\right], 
	\mathbf{X}=\Delta \mathbf{G}_{k}^{\mathrm{H}}.\\
	\end{array}
	\end{aligned}
	\end{equation}
	
	Therefore, by substituting it into (29), we can obtain
\begin{equation}
\begin{aligned}
\left[\begin{array}{ccc}
IN_{k}-\sigma_{k}^{2}-\lambda_{\mathrm{g}, k} M & \widehat{\mathbf{T}}_{k}^{\mathrm{H}} & \mathbf{0}_{1 \times N} \\
\widehat{\mathbf{T}}_{k} & \mathbf{I}_{(K-1)} & \xi_{\mathrm{g}, k} \mathbf{W}_{-k}^{\mathrm{H}} \\
\mathbf{0}_{N \times 1} & \xi_{\mathrm{g}, k} \mathbf{W}_{-k} & \lambda_{\mathrm{g}, k} \mathbf{I}_{N}
\end{array}\right] \succeq \mathbf{0}, 
\end{aligned}
\end{equation}
here $\boldsymbol{\lambda}_{\mathrm{g}}=\left[\lambda_{\mathrm{g}, 1}, \ldots, \lambda_{\mathrm{g}, K}\right]^{\mathrm{T}} \geq 0, \forall k \in \mathcal{K}$
 are slack variables.
Therefore, we can rewrite $\text {(P1')}$ as
\begin{subequations}
	\begin{align}
	\text {(P1'')}
	\min _{\mathbf{W}, \mathbf{e}, \mathbf {IN},\boldsymbol{\tau}_{\mathrm{g}}, \boldsymbol{\lambda}_{\mathrm{g}}}
	& \sum_{k=1}^{K}\left\|\boldsymbol{w}_{k}\right\|^{2}\\
	\text { s.t. } &
\text {(9c),(23b),(30),}
	 \\
	& \boldsymbol{\tau}_{\mathrm{g}} \geq 0, \boldsymbol{\lambda}_{\mathrm{g}} \geq 0.
	\end{align}
\end{subequations}
\subsection{Optimizing Precoder}

Given that $\mathbf{W}$ and $\mathbf{e}$ are intertwined in $\mathbf{J}_{k}$, $\mathbf{j}_{k}$, ${j}_{k}$, $\mathbf{T}_{k}$, and $\widehat{\mathbf{T}}_{k}$, solving problem $\text{(P1'')}$ becomes challenging. To address this, we adopt an iterative approach within AO framework to sequentially optimize $\mathbf{W}$ and $\mathbf{e}$. 
Next, we focus on reformulating the subproblem of $\mathbf{W}$ given a fixed $\mathbf{e}$. This reformulation will enable us to iteratively optimize $\mathbf{W}$ in the AO framework.
\begin{subequations}
	\begin{align}
		\mathbf{W}^{(n+1)}=\arg \min _
{\mathbf{W}, \mathbf {IN},\boldsymbol{\tau}_{\mathrm{g}}, \boldsymbol{\lambda}_{\mathrm{g}}}
&
\sum_{k=1}^{K}\left\|\boldsymbol{w}_{k}\right\|^{2} \\
\text { s.t. } & 
\text {(23b),(30),(31c)} 
	\end{align}
\end{subequations}

Here, we can use the CVX tool~\cite{boyd2004convex} to optimize problem (32) as it is a  semidefinite program (SDP)~\cite{boyd2004convex}.
\subsection{Optimizing Passive Beamforming}

Given $\mathbf{W}$, the problem about $\mathbf{e}$ can be designed as a feasibility-check problem~\cite{wu2019intelligent,9180053,9110587}. 
However, in order to improve the convergence and find a more efficient phase shift solution, we further modify the initial problem $\text{(P1'')}$ into an explicit optimization problem.
Additionally, we use slack variables $\boldsymbol{\beta}=\left[\beta_{1}, \ldots, \beta_{K}\right]^{\mathrm{T}} \ge 0$ to improve the convergence of the optimization problem.
These slack variables represent the residual of the SINR for each user~\cite{wu2019intelligent,9180053}. By incorporating these slack variables, we can effectively manage and minimize interference between users.
Then, (10) is rewritten as
\begin{equation}
\begin{aligned}
\left|\left(\mathbf{h}_{k}^{\mathrm{H}}+\mathbf{e}^{\mathrm{H}} \mathbf{G}_{k}\right) \boldsymbol{w}_{k}\right|^{2} \geq IN_{k}\left(2^{\gamma_{k}/B}-1\right)+\beta_{k}.
\end{aligned}
\end{equation}

Furthermore, the (23b) is rewritten as
\begin{equation}
\begin{aligned}
\left[\begin{array}{cc}
\tau_{\mathrm{g}, k} \mathbf{I}_{M N}+\mathbf{J}_{k} & \mathbf{j}_{k} \\
\mathbf{j}_{k}^{\mathrm{H}} & C_{k}-\beta_{k}
\end{array}\right] \succeq \mathbf{0}.
\end{aligned}
\end{equation}

Then, it is possible to reduce the dimension of the LMI mentioned in equation (30) from $(K+N) \times (K+N)$ to $K \times K$~\cite{9180053}. This reduction allows for a more efficient representation of the LMI by shrinking its size in terms of dimensions while still preserving its essential characteristics.
\begin{equation}
\begin{aligned}
\left[\begin{array}{cc}
IN_{k}-\sigma_{k}^{2}-\lambda_{\mathrm{g}, k} M & \widehat{\mathbf{T}}_{k}^{\mathrm{H}}  \\
\widehat{\mathbf{T}}_{k} & \mathbf{I}_{(K-1)} 
\end{array}\right] \succeq \mathbf{0},
\end{aligned}
\end{equation}

 Thus, we have the problem of $\mathbf{e}$:  
 \begin{subequations}
 	\begin{align}
 	\text {(P1''')}
 	\max _{\mathbf{e}, \mathbf{IN}, \boldsymbol{\beta}, \boldsymbol{\tau}_{\mathrm{g}}, \boldsymbol{\lambda}_{\mathrm{g}}} 
 	& \|\boldsymbol{\beta}\|_{1} \\
 	\text { s.t. } & \text { (9c),(31c),(34),(35)} \\
 	& \boldsymbol{\beta} \geq 0
 	\end{align}
 \end{subequations}

 Due to the constraints (9c), the $\text {(P1''')}$ is non-convex. 
Penalty CCP is adopted to solve this problem~\cite{9180053,9110587}, which can find feasible solutions that meet the unit modulus constraint and QoS constraint.
The constraints (9c) are initially reformulated as $1 \leq \left|e_{m}\right|^{2} \leq 1$ for all $m \in \mathcal{M}$. 
With $e_{m}^{[\iota ]}$ fixed, the constraints are linearized using the expression $|e_{m}^{[\iota ]}|^{2}-2 \operatorname{Re}(e_{m}^{\mathrm{H}} e_{m}^{[\iota ]}) \leq -1$ for all $m \in \mathcal{M}$. 
Therefore, we have 
\begin{subequations}
	\begin{align}
	\max _{\mathbf{e}, \mathbf{d},\mathbf{IN}, \boldsymbol{\beta}, \boldsymbol{\tau}_{\mathrm{g}}, \boldsymbol{\lambda}_{\mathrm{g}}} 
	& \|\boldsymbol{\beta}\|_{1}-\varrho ^{[\iota ]}\|\mathbf{d}\|_{1} \\
	\text { s.t. } & \text {(31c),(34),(35),(36c)} \\
	& \left|e_{m}^{[\iota]}\right|^{2}-2 \operatorname{Re}\left(e_{m}^{\mathrm{H}} e_{m}^{[\iota ]}\right)
	\leq d_{m}-1,  \\
	& \left|e_{m}\right|^{2} \leq 1+d_{M+m}, \forall m \in \mathcal{M}, \\
	& \mathbf{d} \geq 0,
	\end{align}
\end{subequations}
here slack variables are $\mathbf{d}=\left[d_{1}, \ldots, d_{2 M}\right]^{\mathrm{T}}$, and the penalty term $\|\mathbf{d}\|_{1}$ in (37a) is regulated by the regularization factor $\varrho^{[\iota ]}$.
It is worth noting that problem (37) is a SDP problem, overcomed by using existing optimization solvers like CVX~\cite{boyd2004convex}.
We initially design precoder while keeping the reflection beamforming fixed to solve optimization problem.
Next, the $\mathbf{W}$ is held constant, and we address the problem of $\mathbf{e}$
using penalty CCP method. Finally, please refer to Algorithm 1 and Algorithm 2.

\begin{algorithm} [t!]
	\caption{ Alternating Optimization Method}
	\begin{algorithmic}
		\STATE $\textbf{Input:}$  { $\left\{\mathbf{h}_{ k}\right\}$,$\left\{\mathbf{h}_{r, k}\right\}$, $\left\{\mathbf{h}_{b, r}\right\}$ and $\left\{R_{k}^{0}\right\}$ }\\
		\STATE $\textbf{Output:}$ $\left\{c_{f}\right\}$,$\mathbf{W} $ and $\mathbf{e} $ \\
		\STATE $\textbf{Initialization:}	$ 
		{Set convergence tolerance $\epsilon>0$, 
			iteration index $n = 0$ for outer layer, $\mathbf{e}^{(0)}$,  $\mathbf{W}^{(0)}$.} 
			\STATE S1. Find  $\left\{c_{f}\right\} $ by solving (8). 
		\STATE S2. \textbf{repeat}\\
		\STATE S3.\hspace{0.3cm} Update $\mathbf{W}^{(n+1)}$ 
		  , given $\mathbf{e}^{(n)}$ 
		\STATE S4.\hspace{0.3cm} Update $ \mathbf{e}^{(n+1)}$ 
		 , given $\mathbf{W}^{(n+1)}$
		\STATE S5.\hspace{0.3cm} $n \leftarrow n+1 $
		\STATE S6. \textbf{until} 
		 $\left\|\mathbf{W}^{(n+1)}\right\|_{F}^{2}$ converges. 
		 \STATE S7. Return $\left\{c_{f}\right\}$, $\mathbf{W} $ and $\mathbf{e} $ 
	\end{algorithmic}
\end{algorithm}
\begin{algorithm} [t!]
	\begin{algorithmic}
		\caption{ Penalty CCP for passive beamforming
			  }
		\STATE $\textbf{Initialization:}$ {$\mathbf{e}^{[0]}$, $\phi^{[0]}>1$, $\iota =0$.} \\
		\STATE \hspace{0.2cm} \textbf{repeat}\\
		\STATE S1.\hspace{0.3cm} if $ \iota <\iota _{\max } $ then  \\
		\STATE S2.\hspace{0.8cm} Update $\mathbf{e}^{[\iota +1]}$ from (37); \\
		\STATE S3.\hspace{0.8cm}
		$\varrho^{[\iota +1]}=\min \left\{\phi \varrho^{[\iota ]}, \varrho_{\max }\right\}$; \\
		\STATE S4.\hspace{0.8cm} $ \iota =\iota +1 $; \\
		\STATE S5.\hspace{0.3cm} else \\
		\STATE S6.\hspace{0.8cm} Initialize with a new $\mathbf{e}^{[0]}$, 
		 $\phi^{[0]}>1$,
		$\iota =0$ \\
		\STATE S7.\hspace{0.3cm}  end if  \\
		\STATE S8.  \textbf{until} $\|\mathbf{d}\|_{1} \leq \varsigma \text{ ,}\left\|\mathbf{e}^{[\iota ]}-\mathbf{e}^{[\iota -1]}\right\|_{1} \leq \psi  $. \\
		\STATE S9.  Output $\mathbf{e}^{(n+1)}=\mathbf{e}^{[\iota ]}$.
	\end{algorithmic}
\end{algorithm}
{Algorithm development:}
Note that we provide several explanations on Algorithm 2.
a) When the penalty parameter $\varsigma$ is sufficiently low, the constraints in the  (37) can be guaranteed by setting the norm of vector $\mathbf{d}$ as $\|\mathbf{d}\|_{1} \leq \varsigma$.
b) To avoid potential numerical issues, we impose a maximum value $\varrho_{\max}$ to constrain the iteration process. 
As the iteration approaches $\left\|\mathbf{e}^{[\iota ]}-\mathbf{e}^{[\iota -1]}\right\|_{1} \leq \psi$, this feasible solution satisfying $\|\mathbf{d}\|_{1} \leq \varsigma$ may not be found. This maximum limit ensures that the algorithm does not become stuck in such situations.
c) 
 $\left\|\mathbf{e}^{[\iota ]}-\mathbf{e}^{[\iota -1]}\right\|_{1} \leq \psi$ manage the convergence of Algorithm 1. It determines when the iterations can be terminated, achieving the desired solution.
d) In~\cite{9180053,9110587}, a feasible solution of (37) may not necessarily be feasible for (36). To guarantee the feasibility of (36), we set a maximum number of iterations $\varrho_{\max}$. 
If this limit is reached, the iteration process with a new initial point needs to be restarted.

{Complexity Analysis}
First, 
note that higher-level power will cover lower-level power in terms of time complexity.
Let's proceed with analyzing this computational complexity.
For Algorithm 1, content placement problem is addressed in Step 1. By employing KKT-based approach, this problem was solved efficiently with a polynomial-time complexity.
%
 Step 3 to Step 6 are then executed alternately to optimize joint beamforming. 
  The complexity of (32) is 
  $o_{\mathbf{W}}=\mathcal{O}([K(M N+K+N+1)]^{1/2} N K\left[(N K)^{2}+ \right.\\ \left.
  N K^{2}((M N+1)^{2}+ 
  (K+N)^{2})
  + K((M N+1)^{3}+(K+N)^{3})\right])$ 
and that of (37) is 
%
$o_{\mathbf{e}}=\mathcal{O}\left([K(M N+ \right.\\ \left.
1+K)+2 M]^{1 / 2} M \left[M^{2}+
M K\left((M N+1)^{2}+K^{2}\right)+K\left((M N+1)^{3}+K^{3}\right)+M^{2}\right]\right)$.
 Thus, ($o_{\mathbf{W}}+o_{\mathbf{e}}$) is the approximate complexity.

\section{Results} 

\subsection{Parameter Settings}

\begin{table}[ht]
	\centering
	\small
	\caption{{Parameters }}
	\label{cc}
	\setlength{\tabcolsep}{3mm}{
		\begin{tabular}{c<{\centering}c<{\centering}c<{\centering}}
			\toprule[1pt]
			\textbf{Parameter}& \textbf{Value} \\
			\midrule[0.5pt]
			Path loss exponents 
			 {((BS-UE),(BS-IRS),(IRS-UE))}  & 4,2.2,2 \\
			 	Tolerable rate outage probability& 0.05\\
			 	Number of Files $F$ &200\\
			 	Antenna spacing  & half-wavelength\\	
			Noise variance $\sigma^{2}$ & -80 {dBm}\\
			Target rates of all users & $\gamma_{1}, \ldots, \gamma_{K}=\gamma$\\
			System bandwidth $B$ &10 MHz\\
			Local storage size & $100$\\
			Pricing factor $\eta$ & $100$\\
			Stop criterion & $10^{-3}$ \\
			Number of antennas $N$ &6,8,10\\
			Number of reflection units $M$ &6,8,10\\
			\bottomrule[1pt]
	\end{tabular}}
\end{table}

We employ a simulation system configured as illustrated in Table I. 
Moreover, distance dependent path loss models for all channels are described in~\cite{9180053}.
We model small-scale fading as Rician fading model. 
Rician factors are set as follows: $\beta_{BR}=\beta_{Ru}=10$ dB and $\beta_{Bu}=1$ dB. Additionally, the size of the region is controlled by the parameter $\xi_{g,k}$, which represents the radius of the uncertain region in the bounded CSI model in (6).
We reformulate $\xi_{g,k}$ as $\sqrt{0.5 \varepsilon_{g,k}^2 \Phi_{\chi_{2MN}^2}^{-1}(1-\rho)}$, where $\Phi_{\chi_{2MN}^2}^{-1}(\cdot)$ denotes the inverse cumulative distribution function of  Chi-square distribution with $2MN$ degrees of freedom~\cite{chi2017convex}. 
The variance of $\operatorname{vec}(\Delta \mathbf{G}_k)$ is rewritten as $\varepsilon_{g,k}^2=\delta_g^2 \|\operatorname{vec}(\widehat{\mathbf{G}}_k)\|_2^2$.
Here \(\delta_g \in [0,1)\) quantifies the proportionate level of CSI uncertainty.
We consider the following caching strategies and beamforming schemes:

\begin{itemize}
\item {Uniform caching (UC): 
	BS caches content files with same probability and does not depend on their popularity distribution.}
\item {Optimized caching (OC): This is optimized by solving $\mathcal{P}_{2}$.}
\item {Optimized joint beamforming (OJB): Precoder and reflecting beamforming are optimized.
}
\item {Random phase: RIS phase are randomly generated, while precoder is designed by solving (32).}
\end{itemize}

\subsection{Performance Analyses}
Note that evaluation of robustness encompasses multiple test scenarios rather than being restricted to a single experiment.

\emph{1) Impact of the target rate}

\begin{figure}[H]
	\centering
	\subfigure[Transmit power versus target rate]{
		\includegraphics[width=0.45\textwidth]{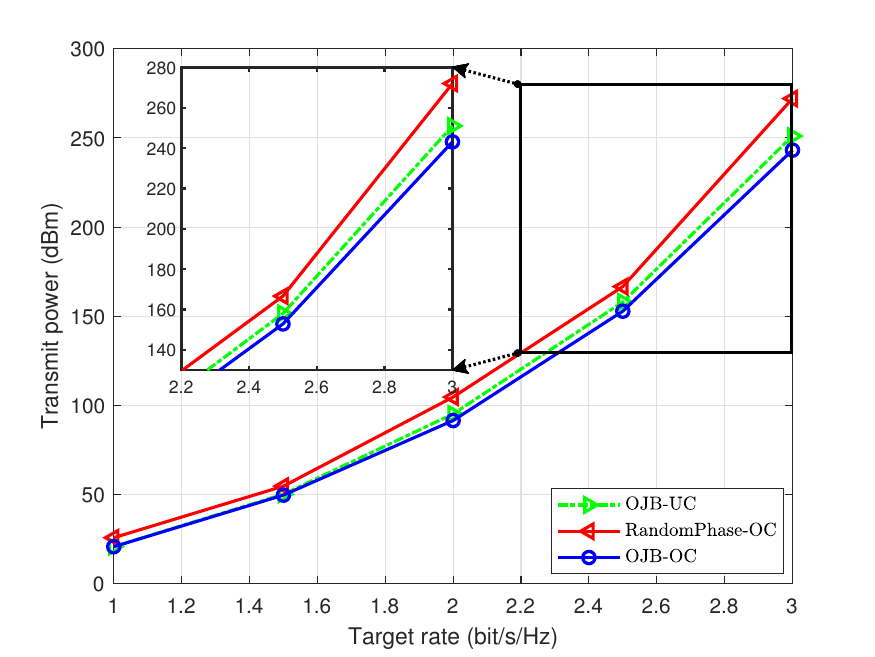}
	}
	\hfill
	\subfigure[Network versus target rate]{
		\includegraphics[width=0.45\textwidth]{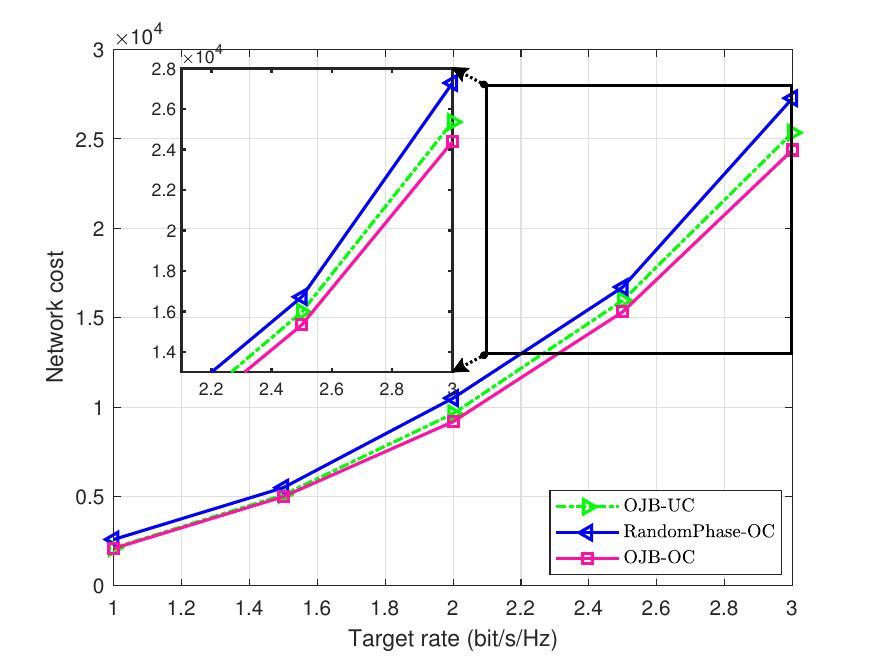}
	}
	\caption{Impact of the target rate}
	\label{fig_4}
\end{figure}

{Here, we set $\delta_{\mathrm{g}}= 0.01$, $K =3$, $ \varepsilon=1 $}. Fig. \ref{fig_4}.(a) illustrate transmission power under different target rate.
    Firstly, as the target rate increases, the minimum transmission power of the scheme with the OJB-UC, RandomPhase-OC and OJB-OC (i.e., our proposed scheme) increases. It should be noted that compared to other benchmark schemes, the proposed scheme has an advantage in power consumption. 
    Specifically, for high target rate (i.e., 3bit/s/Hz), the proposed solution saves approximately $11.89\%$ of power consumption compared to RandomPhase-OC and approximately $3.24\%$ compared to OJB-UC. 
    Unlike equal probability caching, which does not account for content popularity, optimizing the caching content placement can leverage the benefits of skewed content popularity as well as user request distribution.
    Noted that the proposed method simultaneously optimizing content placement and joint beamforming achieves the lowest power cost, as the quality of offloading links is enhanced by optimally configuring phase shifts.
    
Fig. \ref{fig_4}.(b) represents the network cost versus target rates of the users $K=3$.
    {Then, we set $\delta_{\mathrm{g}}= 0.01$, $ \varepsilon=1 $}.
    Then, as the target rate increases, the minimum network cost of the all schemes increases. Furthermore, it can be seen that among all the schemes, the proposed OJB-OC has the lowest network cost. 
    This is because the reflected signals have been effectively combined towards the receivers. 
    In contrast, the proposed hybrid beamforming design significantly reduces the network cost.
    Additionally, the proposed caching strategy reduces the burden on the backhaul link, further improving network costs.

{\emph{2) Impact of IRS elements and number of transmitting antennas}}

\begin{figure}[t!]
	\centering
	\subfigure[Transmit power versus target rate 
	with $K=3$]{
		\includegraphics[width=0.45\textwidth]{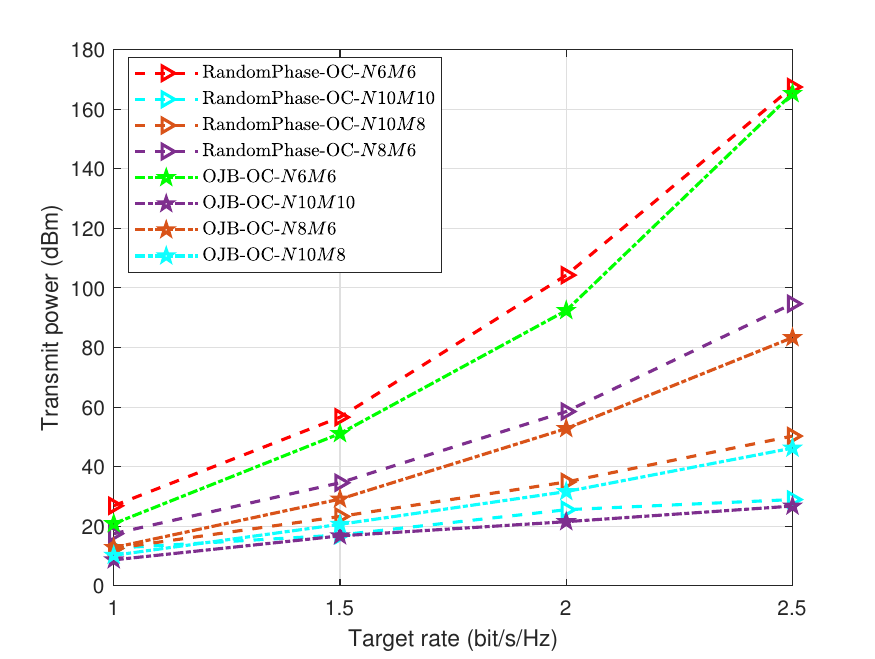}
	}
	\hfill
	\subfigure[Network versus target rate with $K=3$]{
		\includegraphics[width=0.45\textwidth]{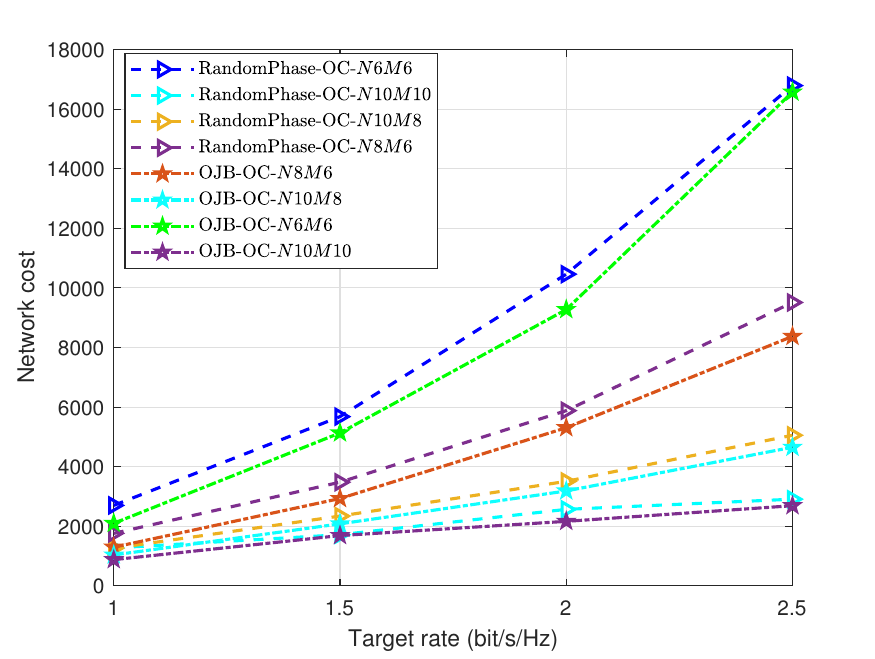}
	}
	\caption{Impact of IRS elements and number of transmitting antennas}
	\label{fig_6}
\end{figure}

The impact of i.i.d. Rician model on the minimum transmit power and network cost while varying the target rate and the number of elements and transmitting antennas in the Fig. \ref{fig_6}. Next, we set $K =3$, $\delta_{\mathrm{g}}= 0.01$, $ \varepsilon=1 $,
$ N=6,8,10 $ and $ M=6,8,10 $.
The tolerable rate outage probability is set to 0.05.
Then, the transmission power increases with the increase of
target rates in all solutions. Afterwards, compared to the RandomPhase-OC, it indicates that transmission power is reduced with more RIS units for the proposed schemes, especially outstanding in high target rate.
This is because larger $M$ obtain more favorable line of sight
link gain, which decrease transmission power.
Then, increasing the number of transmitting antennas helps BS achieve better transmission gain.

Subsequently, when $N=10$ and $M=10$, the proposed scheme has certain advantages compared to the baseline scheme.
It should be noted that the above phenomenon also exists in network costs.
Meanwhile, the reasonable selection of reflective elements and transmitting antennas  for the content placement and active/passive beamforming co-design provides engineering insights for practical deployment.

\emph{3) Impact of the CSI uncertainty}

\begin{figure}[H]
	\centering
	\subfigure[{Transmit power versus target rate (different CSI error)}]{
		\includegraphics[width=0.45\textwidth]{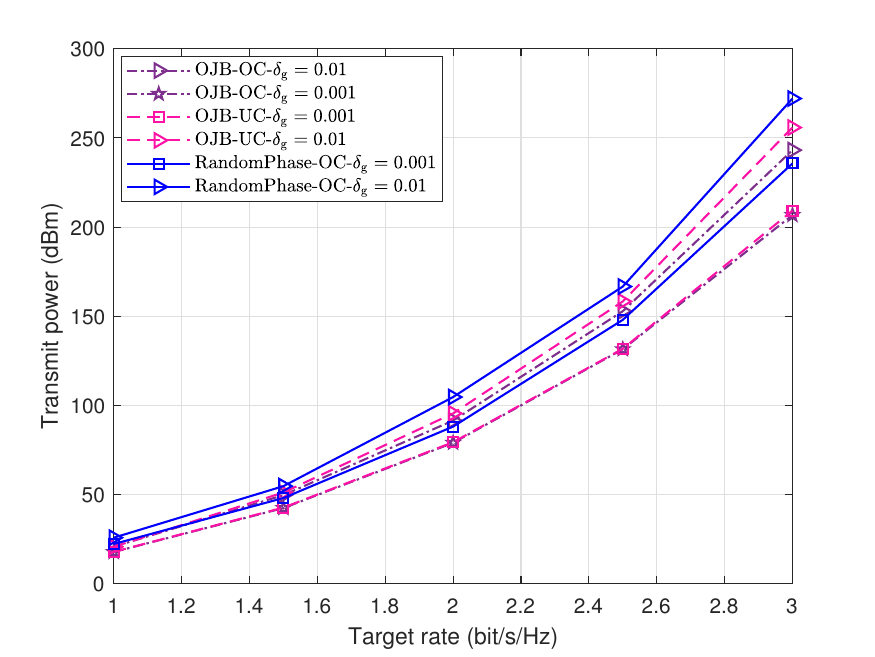}
	}
	\hfill
	\subfigure[{Network cost versus target rate (different CSI error)}]{
		\includegraphics[width=0.45\textwidth]{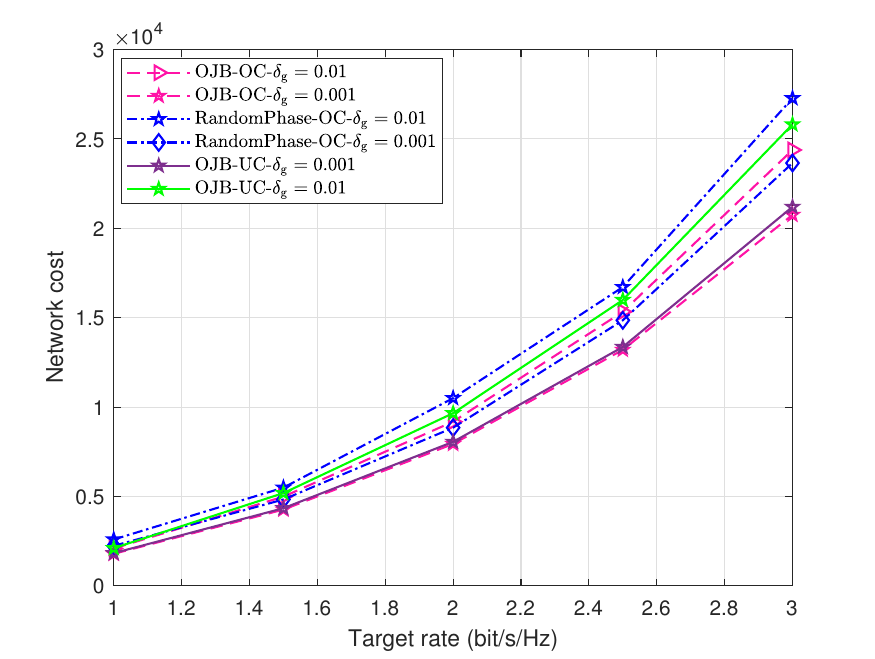}
	}
\caption{Impact of different CSI uncertainty}
 \label{fig_8}
\end{figure}

The impact of i.i.d. Rician model on the minimum transmit power and network cost while varying the target rate and the CSI uncertainty in the Fig. \ref{fig_8}. Next, we set $K =3$, $\delta_{\mathrm{g}}= 0.01,0.001$, 
and $ \varepsilon=1 $.
Then, as the target rate increases, the minimum transmit power increases with the all schemes. An intriguing observation is evident from Fig. \ref{fig_8}.(a). The minimum transmission power in the OJB-OC is lower than that in other benchmark schemes due to the combined influence of two factors: the level of CSI uncertainties and the target rates. Then, for Fig. \ref{fig_8}.(a),  the transmit power value under the OJB-OC ($\delta_{\mathrm{g}}= 0.001$) is the best, followed by {OJB-URC} ($\delta_{\mathrm{g}}= 0.001$),
 the worst is {RandomPhase-OC} ($\delta_{\mathrm{g}}= 0.01$). 
In addition, it is imperative to thoroughly investigate the impact of channel uncertainty, as they play a crucial role in system performance.
This is because worst-case optimization represents the most conservative robust design, demanding more power to guarantee that the achievable rate for each user fulfills  target rate requirement under the realization of CSI error. This phenomenon is witnessed in all schemes. 

The relationship between network cost and target rate is shown in the Fig. \ref{fig_8}.(b).
Then, the network cost of all solutions increases as the target rate increases. More specifically, the OJB-OC exhibit distinct advantages in terms of the network cost, as evidenced by the performance depicted in the Fig. \ref{fig_8}.(b).
The joint design of content placement and active/passive beamforming has been optimized to reduce backhaul burden and improve transmission gain, which can conserve network costs.
Notably, the Fig. \ref{fig_8}.(b) highlights that the network costs performance is considerably sensitive to the magnitude of the CSI errors, emphasizing the need for precise CSI considerations. This will be further discussed in next work.

In summary, the objective of this article is to explore the influence of 
 factors including imperfect CSI, content placement optimization, joint beamforming on system performance of the IRS enables edge caching for IoT.
The findings aim to provide valuable insights for the design and deployment of wireless networks via IRS in various practical scenarios and specific situations. 

\subsection{Discussion}
\emph{Remark}: 
Nonetheless, when it comes to the beamforming design and the corresponding methods, certain aspects warrant further discussion. The key insights from this analysis can be summarized as follows.
\begin{itemize}
\item {The current plan considers a caching strategy (i.e., files are fully cached at the transmitter). 
	The disadvantage is that as file diversity increases, the service distance increases, making it difficult to fully leverage the advantages of file diversity. A segmentation based encoding caching strategy can be considered to improve performance. 
While it is assumed that the popularity of files is known, this popularity needs to be predicted in actual networks, which will be addressed in subsequent work.
}
\item {
	Then, the proposed model did not take into account hardware damage of transceiver.
	This aspect will be investigated in future work. }	
	\item { Moreover, the Zipf model can only analyze static content and is difficult to handle time-varying information. Therefore, we will consider an adaptive caching scheme in future work.}
\item {
	Final, this article only focuses on the co-design under the ideal phase shift model with constant modulus constraints and continuous phase. 
	This assumption may impose certain limitations on the practicality of the system. Next, we consider practical phase shift models~\cite{abeywickrama2020intelligent,saglam2022deep} in our future work.
	}
\end{itemize}

\section{Conclusions}
First, we consider content placement and active/passive beamforming co-design problem for IRS-based MIMO system under imperfect CSI.
However, due to the non-convex unit-modulus constraints and worst-case QoS, as well as the coupling of variables, the optimization problem is non-convex and cannot be tackled straightforwardly.
After decouplingcontent placement subproblem from beamforming design, constraints are linearly approximated to solvable equivalent constraintsby using S-procedure, Schur complement, Sign-definiteness lemma.
Meanwhile, content placement problem can be addressed by applying the KKT optimality conditions.
Afterwards, we proposed an alternating optimization framework to address precoder and reflection beamforming. We then employ the penalty CCP method and CVX techniques under this framework to obtain suboptimal solutions.
{Numerical simulation reveals that OJB-OC effectively reduces transmission power and system network costs, and is superior to traditional solutions.
Results indicated that caching content placement and beamforming co-design
could more effectively utilize the limited MEC storage resources of the system
while ensuring the reliability of transmission links.
}	
%

\IEEEtriggeratref{53} %

%
%
%
%

\end{document}